\documentclass[prx, reprint,
 superscriptaddress,
 amsmath,amssymb,
 aps,
 tikz,border=0pt,
]{revtex4-2}
\usepackage[utf8]{inputenc}
\usepackage{braket}
\usepackage{graphicx}
\usepackage{dcolumn}
\usepackage{bm}
\usepackage{booktabs}
\usepackage{xcolor}
\begin{document}
\title{One-dimensional electronics with edge states in two-dimensional altermagnets}
\author{Shibo Fang}
\affiliation{Science, Mathematics and Technology (SMT) Cluster, Singapore University of Technology and Design, Singapore 487372}
\author{Zongmeng Yang}
\affiliation{Science, Mathematics and Technology (SMT) Cluster, Singapore University of Technology and Design, Singapore 487372}
\affiliation{State Key Laboratory for Mesoscopic Physics and School of Physics, Peking University, Beijing 100871, P. R. China}
\author{Jianhua Wang}
\affiliation{Science, Mathematics and Technology (SMT) Cluster, Singapore University of Technology and Design, Singapore 487372}
\affiliation{Institute for Superconducting and Electronic Materials, Faculty of Engineering and Information Sciences, University of Wollongong, Wollongong 2500, Australia}
\author{Xingyue Yang}
\affiliation{Science, Mathematics and Technology (SMT) Cluster, Singapore University of Technology and Design, Singapore 487372}
\affiliation{State Key Laboratory for Mesoscopic Physics and School of Physics, Peking University, Beijing 100871, P. R. China}
\author{Jing Lu}
\affiliation{State Key Laboratory for Mesoscopic Physics and School of Physics, Peking University, Beijing 100871, P. R. China}
\author{Ching Hua Lee}
\affiliation{Department of Physics, National University of Singapore, Singapore 117551, Singapore}
\author{Xiaotian Wang}
\email{xiaotianw@uow.edu.au}
\affiliation{Institute for Superconducting and Electronic Materials, Faculty of Engineering and Information Sciences, University of Wollongong, Wollongong 2500, Australia}
\author{Yee Sin Ang}
\email{yeesin\_ang@sutd.edu.sg}
\affiliation{Science, Mathematics and Technology (SMT) Cluster, Singapore University of Technology and Design, Singapore 487372}

\begin{abstract}
 The coupling between real-space inhomogeneity coordinates and spin ($\boldsymbol{r}$--$\boldsymbol{s}$) provides an alternative route to achieve efficient spin manipulation in spintronics beyond the conventional momentum-spin ($\boldsymbol{k}$--$\boldsymbol{s}$) coupling paradigm. Here we demonstrate an unexpected manifestation of one-dimensional (1D) $\boldsymbol{r}$--$\boldsymbol{s}$ coupling via floating edge states in two-dimensional (2D) altermagnets with electric-field and gate-doping tunability. The 1D edge-spin $\boldsymbol{r}$--$\boldsymbol{s}$ coupling ensures that carrier transport is exclusively carried by the edge states with quantized spin conductance, giving rise to an unconventional edge tunnel magnetoresistance (edge-TMR) effect that can be switched \textit{on} or \textit{off}. As a proof of concept, we computationally design an altermagnetic edge tunnel junction (edge-MTJ) based on a Cr$_2$Se$_2$O monolayer to demonstrate its edge transport and controllability via the Néel order or an electric field. Our findings propose a general prototype altermagnetic device for next-generation low-dimensional spintronics.
\end{abstract}

\maketitle
\section{Introduction}
Spintronics harnesses spin degrees of freedom in solid-state systems for information processing \cite{vzutic2004spintronics,baltz2018antiferromagnetic}. The key operational principle of spintronics lies in the coupling between spin $\boldsymbol{s}$ and other degrees of freedom \cite{guo2025spin,pulizzi2012spintronics}, so as to achieve efficient spin manipulation. In materials with periodic boundary conditions, the crystal momentum $\boldsymbol{k}$ is a good quantum number, and the coupling between $\boldsymbol{k}$ and $\boldsymbol{s}$ gives rise to rich physical properties beneficial for device applications \cite{shao2024antiferromagnetic,vsmejkal2022giant,vsmejkal2022emerging}. The  $\boldsymbol{k}$-$\boldsymbol{s}$ coupling is a key enabler of nonvolatile memory functionalities in magnetic tunnel junctions (MTJs) \cite{wu2023high,tsymbal2003spin,zhang2021recent}, driving the recent emergence of a diverse range of MTJ devices based on various $\boldsymbol{k}$-$\boldsymbol{s}$-coupled systems such as altermagnets \cite{wang2025pentagonal,noh2025tunneling,chi2025anisotropic}, ferroelectric altermagnets \cite{gu2025ferroelectric,duan2025antiferroelectric,sun2025proposing,cao2024designingspindrivenmultiferroicsaltermagnets}, noncollinear antiferromagnets \cite{rimmler2025non,dong2022tunneling}, ferroelectric systems with strong relativistic effects \cite{tao2025ferroelectric,huang2025two}, and spin-valley-mismatched ferromagnetic metals \cite{yan2025giant}. 

In nanoscale devices, the breaking of translational symmetry within the device is more prominent due to the small length scales, and the wave vector $\boldsymbol{k}$ is no longer a suitable quantum number for describing wave functions \cite{yia1982complex}. Instead, the spatial coordinate $\boldsymbol{r}$ becomes an alternative label of the wave function. For instance, in the 2D layered van der Waals systems, the control of interlayer degrees of freedom gives rise to a rich variety of physical mechanisms \cite{lee2018electromagnetic}, such as the layer Hall effect \cite{gao2021layer} and layer-spin coupling \cite{fang2024light}. In the zero-dimensional (0D) systems such as quantum dots \cite{barry2020sensitivity,gong2024hidden,han2024cornertronics} and molecular systems \cite{wang2025critical}, the wave function is characterized by position indices $\boldsymbol{r}$ and energy levels $\textit{n}$, instead of $\boldsymbol{k}$. Moreover, the mismatch between magnetic sublattices can also be interpreted as a manifestation of spintronic phenomena arising from spatial confinement \cite{yang2025interface}. When considering spintronics in spatially confined systems, it is thus natural to introduce the coupling between $\boldsymbol{r}$ and $\boldsymbol{s}$, i.e., $\boldsymbol{r}$-$\boldsymbol{s}$ coupling, for spintronic device applications. 
Physical phenomena related to $\boldsymbol{r}$-$\boldsymbol{s}$ coupling have been actively studied in recent years, covering topological insulator \cite{kane2005z,kane2005quantum}, hidden spin polarization \cite{zhang2014hidden,chen2022role}, Néel spin current \cite{shao2023neel,fang2024light}, altermagnetism \cite{liu2025different, fender2025altermagnetism}, layertronics \cite{yao2025switching,gong2018electrically}, and cornertronics \cite{han2024cornertronics,ren2020engineering}. Contrary to the well-studied $\boldsymbol{k}$-$\boldsymbol{s}$-coupled systems, $\boldsymbol{r}$-$\boldsymbol{s}$ coupling spintronics remains largely unexplored \cite{shao2024antiferromagnetic}.

Altermagnets constitute an emerging magnetic subfamily with collinear antiferromagnetic spin alignment and spin-split electronic structures \cite{bai2024altermagnetism,liu2025different,guo2025spin,fu2025allelectricallycontrolledspintronicsaltermagnetic,fu2025lightinducedfloquetspintripletcooper,fu2025floquetengineeringspintriplet,ang2023altermagneticschottkycontact,ma2021multifunctional,hu2025catalog,zhou2025manipulation}. Their unusual nonrelativistic $\boldsymbol{k}$--$\boldsymbol{s}$ coupling endows them with strong potential in antiferromagnetic spintronic applications \cite{baltz2018antiferromagnetic}, particularly as \emph{altermagnetic tunnel junctions} (AMTJ) \cite{shao2024antiferromagnetic,yang2025unconventional,vsmejkal2022giant,wang2025pentagonal}. The mechanism of AMTJs relies on the switching achieved through the (mis)matching of the $\boldsymbol{k}$--$\boldsymbol{s}$ coupling wavefunctions between the two altermagnetic electrodes. From a real-space perspective, the unconventional $\boldsymbol{r}$--$\boldsymbol{s}$ coupling \cite{fender2025altermagnetism} arises when the intrinsic symmetry between the two magnetic sublattices is externally broken, which can give rise to peculiar 0D spin-dependent topological corner states \cite{wang2025pentagonal} and 2D Néel-order-dependent layertronics \cite{zhang2024predictable,peng2025all}. 
The $\boldsymbol{r}$--$\boldsymbol{s}$ coupling in altermagnets is thus expected to offer a platform for the hunt of exotic electron and spin transport phenomena.

\begin{figure}[t]
\centering
\includegraphics[width=\linewidth]{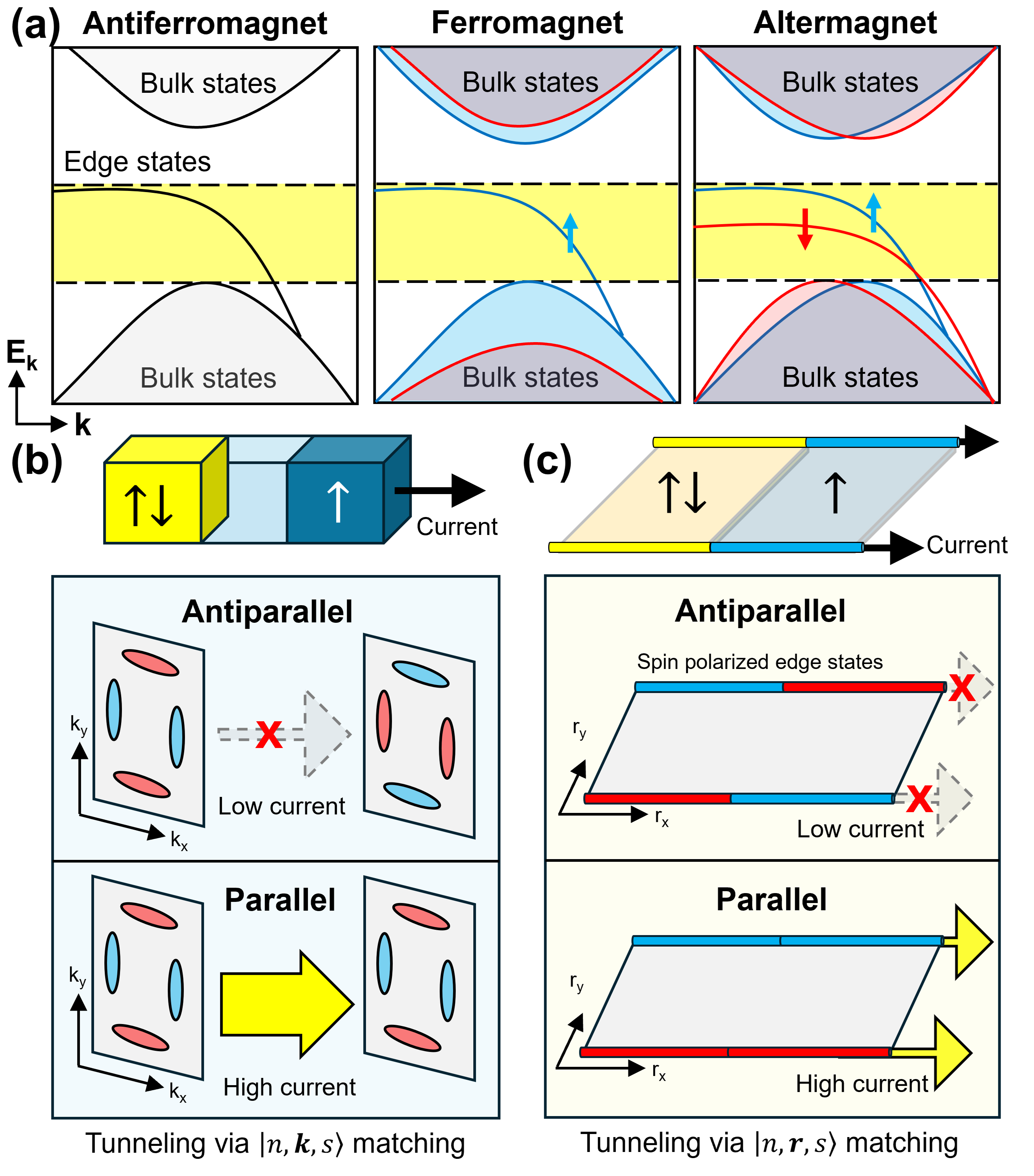} 
\caption{\label{fig1} Characteristics of edge states in altermagnets and the mechanism of the associated edge-MTJ. \textbf{(a)} Schematic illustration of bulk bands and edge states in antiferromagnets, ferromagnets, and altermagnets. \textbf{(b)} Conventional altermagnetic tunnel junctions based on $\boldsymbol{k}$-$\boldsymbol{s}$ coupling. \textbf{(c)} Altermagnetic edge-MTJ based on $\boldsymbol{r}$-$\boldsymbol{s}$ coupling. Here $n$, $k$, $r$, $s$ refer to energy level, momentum, spatial coordinate, and spin, respectively. The red and blue colors in the figure represent different spins. For altermagnets, the spin-split edge states shown here are schematic for suitable ribbon orientations and edge terminations; other terminations may remain spin degenerate due to residual symmetry.}
\end{figure}

In this work, we demonstrate an electrically tunable 1D $r$--$s$ coupling by harnessing the floating edge states of 2D altermagnets. Different from the ferromagnetic and antiferromagnetic systems, these edge states in altermagnets are intrinsically spin-split and are floating between the bulk states [Fig.~\ref{fig1}(a)]. Unlike the conventional AMTJ that relies on $\boldsymbol{k}$--$\boldsymbol{s}$ matching in altermagnetic electrodes [Fig.~\ref{fig1}(b)], we propose a prototype 1D altermagnetic edge tunnel junction (edge-MTJ) exhibiting an unconventional edge tunnel magnetoresistance (edge-TMR) effect (Fig.~\ref{fig1}(c)). Here, the resistance contrast is defined in a manner similar to tunneling magnetoresistance, but it originates from edge-state transport instead of conventional quantum tunneling through a barrier. The electric current in edge-MTJ is entirely carried by edge states, exhibiting ideal single-spin quantum conductance. By tuning the Néel order or applying an external electric field, the electric current can be effectively switched from \textit{on} to \textit{off}, enabling integrated memory and logic functionality. In particular, electrostatic doping can be employed to make the edge-MTJ host only a single spin-polarized conductance channel, a phenomenon that is unique to altermagnets. The edge-MTJ is constructed from a homogeneous material, thereby avoiding common blocking effects caused by the antiferromagnetic interface \cite{yang2025interface}. The edge-counterpart of the MTJ proposed here, namely the edge-MTJ, offers a gate-programmable platform that electrically controls the floating 1D edge states in altermagnets, thereby activating a gate-induced single-spin transport channel and enabling an electrically tunable $\boldsymbol{r}$-$\boldsymbol{s}$ coupling for high-density spintronic memory. 

\section{Results}
\subsection{1D $r$--$s$ coupling in altermagnets}
We consider the floating edge states in 2D second-order topological insulators, which originate from a polarization shift between the Wannier charge centers and the atomic positions \cite{bradlyn2017topological,lee2015free,liu2020topological,xie2021higher,cualuguaru2019higher}. The floating surface state is a low-bandwidth electronic state that appears at the surface of obstructed atomic insulators, residing between the bulk valence and conduction bands and crossing the Fermi level independently of electron occupation \cite{liu2024observation,lee2014lattice}. These floating edge states are fully localized at the boundary and, by directly crossing the Fermi level, contribute to transport far more significantly than the bulk. In altermagnets, we find that the spins of these edge states are directly locked to the Néel order. We refer to applications based on these edge states as 1D $\boldsymbol{r}$-$\boldsymbol{s}$ coupling. The motivation for investigating such edge states lies in their unique origin: their existence does not require spin–orbit coupling-induced band inversion, and they can exist in systems with large band gaps while remaining energetically isolated from the bulk states \cite{liu2024observation,zhu2018quasiparticle}. Moreover, these edge states exhibit high tunability and can be effectively modulated by external fields.

In conventional antiferromagnets, the two magnetic sublattices share identical chemical environments, resulting in symmetric edge-state dispersions for nanoribbons with opposite spin orientations \cite{fu2025building}. In contrast, in altermagnets, the two spin sublattices experience distinct chemical bond orientations, which leads to a splitting in the dispersion relations of their respective edge states \cite{fender2025altermagnetism,song2025altermagnets,chatterjee2025interplay}. This phenomenon can also be regarded as a peculiar manifestation of \emph{1D altermagnetism} in which the 1D electronic bands exhibit spin-contrasting dispersion. Our first-principles calculations demonstrate that the switching can be achieved through the spin matching or mismatching of these floating edge states, which illustrates the application of 1D $\boldsymbol{r}$-$\boldsymbol{s}$ coupling. 

We note that spin-polarized edge states are not unique to altermagnets. In conventional antiferromagnetic nanoribbons, such states may also arise when $\mathcal{PT}$ symmetry is broken by specific edge terminations. However, in that case the spin splitting originates mainly from boundary symmetry breaking, rather than from an intrinsically spin-split bulk electronic structure. By contrast, in altermagnets, the edge-state spin splitting is not merely a boundary-induced effect, but is tied to the underlying bulk altermagnetic electronic structure, and may therefore be more pronounced than in $\mathcal{PT}$-broken antiferromagnetic nanoribbons. This makes altermagnetic edge states a more effective basis for the spin-dependent transport and switching behavior discussed here.
 
\subsection{Edge-MTJ}
Here we propose the concept of the edge-MTJ, a tunneling junction exhibiting switching effect via the spin-matching and mismatching of edge states at the electrodes. As shown in Fig. 1(c), the edge-MTJ is based on a two-terminal device configuration with electrons transversally along the edge of the 2D altermagnets. Switching between high- and low-resistance states can be achieved by changing the N\'eel order on one end of device while keeping that of the other end of the device fixed.

Here we briefly introduce the mechanism of the edge-MTJ. The quantum ballistic of a two-terminal device \cite{quhe2021sub,ding2021two,yang2021layer,fang2023polarization,chen2025self,fang2024tunable} is described by the Landauer–Büttiker formalism \cite{datta1992exclusion,meir1992landauer}:

\begin{equation}
I = \frac{2e}{h} \int dE \, T(E) [f_L(E) - f_R(E)]
\end{equation}

where $E$ is energy, $T(E)$ denotes the transition probability between the left and right electrodes, $f_{L/R}(E)$ represents the Fermi-Dirac distribution function in the left/right electrodes. Here, $T(E)$ is determined by the properties of the left and right electrodes as well as their coupling to the central region. In the absence of coupling between the central region and the electrodes, the charge transport through the central region is prohibited. Under such conditions, electrons can transfer between the left and right electrodes via quantum tunneling, which characterizes a tunneling junction. The transport formula of the system reduces to  the Bardeen model, where the current depends on the tunneling matrix elements between electronic states and the density of states of the two electrodes, in a form analogous to Fermi's golden rule \cite{bardeen1961tunnelling,micklitz2022emergence}:

\begin{align}
I &= \frac{2\pi e}{\hbar} \sum_{\mu,\nu} \left[ f_L(E_\mu) - f_R(E_\nu) \right] |M_{\mu\nu}|^2 \, \delta(E_\mu - E_\nu) \notag\\
&= \frac{2\pi e}{\hbar} \int dE\, \rho_L(E) \rho_R(E) |M(E)|^2 \left[ f_L(E) - f_R(E) \right].
\end{align}

where $\mu, \nu$ are the quantum states in the left and right electrodes, respectively, and $M_{\mu\nu}$ is the tunneling matrix element between states $\mu$ and $\nu$. $M(E)$ is the effective tunneling matrix element at energy $E$, and $\rho_{L/R}(E)$ represents the electronic density of states in the left/right electrodes. 

The resistance switching effect of quantum devices arises from the modifications of the transition probability by changing the magnetic ordering between the two electrodes such as the cases of MTJ and AMTJ \cite{shao2024antiferromagnetic}, or by modulating the ferrolelectric polarization such as the case of ferroelectric tunnel junction \cite{tsymbal2006tunneling}. In edge-MTJs, the resistance switching is modulated by switching the Néel order of the edge states, and can be further modulated using an external lateral electric field. When the edge-MTJ is in current-ON state (referred to as the parallel state, or P state), both the initial and final quantum state are identically in $\ket{r,s}$. In contrast, when the device is switched into the current-OFF state (the antiparallel state, or AP state), the final state becomes spin-opposite to the initial state, i.e. $\ket{r,-s}$. The matching or mismatch of $\boldsymbol{s}$, thus, enables current ON/OFF switching based on the edge-TMR conditions of (Fig. 1(c) is used for a schematic, while Fig. 3(a) and (b) serve as actual devices):

\begin{align}
|r,s\rangle &\rightarrow |r,s\rangle \quad \text{(P state)} \\
|r,s\rangle &\nrightarrow |r,-s\rangle \quad \text{(AP state)}
\end{align}

\begin{figure}[t]
\includegraphics[width=\linewidth]{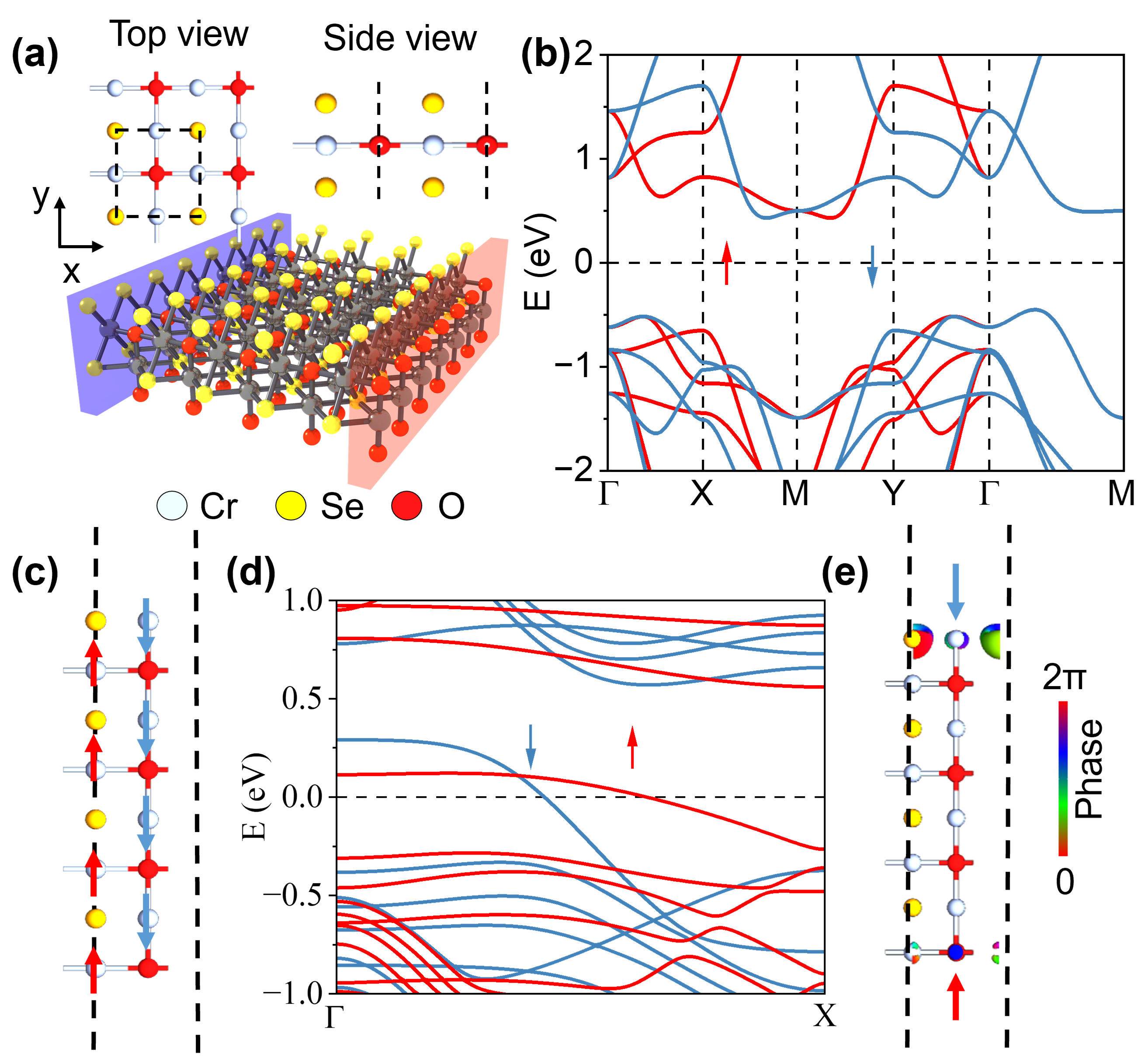}
\caption{\label{fig2} Edge states in 2D altermagnetic Cr$_2$Se$_2$O nanoribbon.\textbf{(a)} Geometry of monolayer (ML) Cr$_2$Se$_2$O and schematic illustration of its nanoribbon.\textbf{(b)} Band structure of ML Cr$_2$Se$_2$O, where the red and blue arrows indicate spin-up and spin-down, respectively.\textbf{(c)} Geometry of the Cr$_2$Se$_2$O nanoribbon with edges of opposite spins and \textbf{(d)} its band structure. \textbf{(e)} Bloch states of \textbf{(d)} in Fermi surface. The color scale indicates the phase of the Bloch wave. The Bloch states at the upper and lower edges are spin-down and spin-up, respectively, reflecting the characteristic of 1D
$\boldsymbol{r}$-$\boldsymbol{s}$ coupling.}
\end{figure}

\subsection{Material realization}
By investigating a variety of 2D altermagnetic nanoribbons, we find that the floating edge states are prevalent in several mirror real Chern insulators, a subclass of second-order topological insulators, including Cr$_2$Se$_2$O, V$_2$Se$_2$O, Cr$_2$S$_2$, Cr$_2$Cl$_2$, and Cr$_2$Br$_2$ (Figs. S1-4) \cite{SupplementalMaterial}. Here, we take monolayer Cr$_2$Se$_2$O as an example to illustrate the properties of the edge-MTJ (Fig.~\ref{fig2}(a)). The red and blue arrows and lines represent spin-up and spin-down states in Fig. 2, respectively. Cr$_2$Se$_2$O belongs to the ${P}^{-1}{4/}^{1}{m}^{1}{m}^{-1}m^{\infty m}1$ spin space group \cite{chen2024enumeration}, with a lattice constant of 4.02~\AA{} and band gap of 1.12 eV (Fig.~\ref{fig2}(b)) \cite{gong2024hidden,yang2025unconventional}.  It exhibits spin degeneracy along $\Gamma$–M, whereas spin splitting occurs along $\Gamma$–X–M–Y–$\Gamma$, indicating its altermagnetic features. The similar layered compounds with Cr$_2$Se$_2$O such as KV$_2$Se$_2$O \cite{jiang2025metallic} and Rb$_{1-\delta}$V$_2$Te$_2$O \cite{zhang2025crystal} have been demonstrated to exhibit altermagnetism in recent Spin- and angle-resolved photoemission spectroscopy (SARPES) experiments. A closely related layered material, V$_2$Se$_2$O, can be exfoliated from its bulk form into a 2D structure \cite{lin2018structure}, suggesting that 2D Cr$_2$Se$_2$O may also be experimentally attainable.

We consider a Cr$_2$Se$_2$O nanoribbon with a width of 4 unit cells and type 1 ordering at the edges as an example (Fig.~\ref{fig2}(c)). Unlike the bulk band structure, the nanoribbons of Cr$_2$Se$_2$O exhibit two spin-split bands crossing the Fermi level, while the other bands retain the semiconducting character of the bulk phase (Fig.~\ref{fig2}(d)). We calculated the real-space projections of the two Bloch states that cross the Fermi level. The results show that both Bloch states are localized at the edges, with their spins aligned with the corresponding Cr atoms at the edges (Fig.~\ref{fig2}(e)).  We also calculate the evolution of such edge states under different nanoribbon widths (Fig. S5) \cite{SupplementalMaterial}. The results show that these edge states exist across different ribbon widths. Using the surface Green’s function method, our results show that these edge states persist even in the limit of infinitely wide nanoribbons (Fig. S6) \cite{SupplementalMaterial}. At present, we focus on vertical edges obtained by cutting along the $y$ direction. If we consider horizontal edges cut along the $x$ direction, the $\mathcal{TC}_4$ symmetry of $\mathrm{Cr_2Se_2O}$ dictates that the spin splitting of the edge-state bands is exactly reversed (Fig.~S7) \cite{SupplementalMaterial}.

It is worth noting that the atomic termination of a nanoribbon can have a pronounced impact on its properties \cite{wang2021graphene,wang2021towards,jiang2025signatures}. This is also the case for the altermagnetic $\mathrm{Cr_2Se_2O}$ nanoribbons considered here. We find that nanoribbons with different edge terminations and cutting directions exhibit distinct band dispersions and spin-splitting characteristics, yet they all host floating edge states [Figs. S8-S9]. Notably, the edge states in nanoribbons cut along the [100] direction are spin split, whereas those along the [110] direction remain spin degenerate. This behavior is dictated by whether the magnetic atoms contributing to the edge states carry the same spin polarization, or instead consist of opposite spins that compensate each other.

\begin{figure}[t]
\centering
\includegraphics[width=\linewidth]{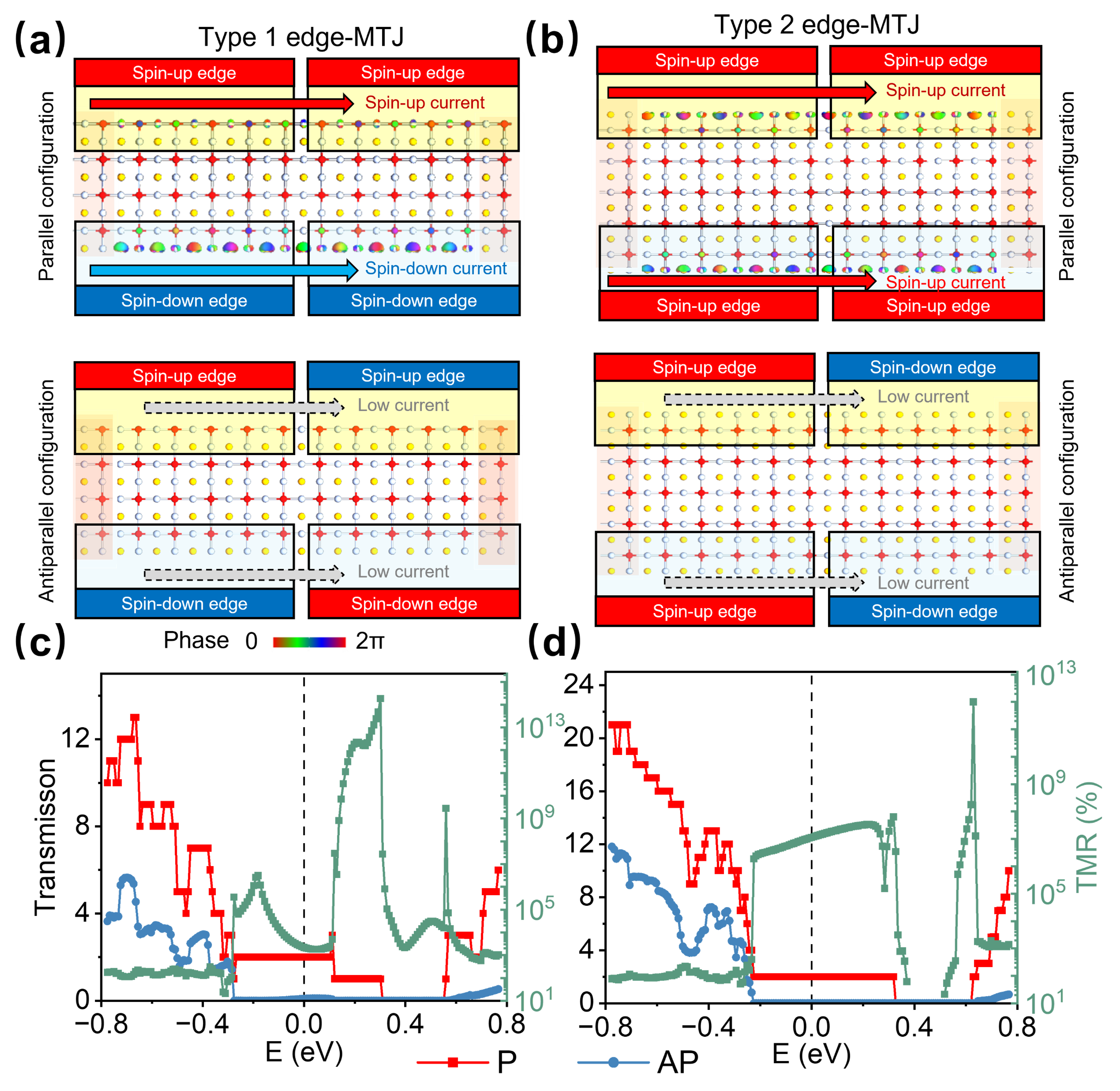}
\caption{\label{fig3} Schematic diagram of the Cr$_2$Se$_2$O edge-MTJ. \textbf{(a)} and \textbf{(b)} illustrate the mechanisms of edge-MTJ with opposite-spin (type 1) and same-spin (type 2) edge, respectively. The color scale indicates the phase of the Bloch wave. \textbf{(c)} and \textbf{(d)} are the transmission spectra and edge-TMR corresponding to the configurations in \textbf{(a)} and \textbf{(b)}, respectively. In the two configurations, the transmission spectrum in the P state exhibits pronounced quantum conductance at the Fermi level, whereas this quantum conductance is suppressed in the AP state.}
\end{figure}

The transport properties of Cr$_2$Se$_2$O nanoribbon are investigated using \textit{ab initio} quantum transport simulation (Fig.~\ref{fig3}) \cite{smidstrup2019quantumatk}. Two types of nanoribbon edge configurations are considered: (i) opposite-spin [type 1, Fig.~\ref{fig3}(a)] and (ii) same-spin [type 2, Fig.~\ref{fig3}(b)] edges (see Fig. S10 \cite{SupplementalMaterial}). Both device structures are cleaved along the $[100]$ crystallographic direction of Cr$_2$Se$_2$O. In the type 1 device, the two edges are terminated by Cr–O and Cr–Se atoms, respectively, whereas in the type 2 device both edges are terminated by Cr–Se atoms. The device states with $\boldsymbol{k}$-$\boldsymbol{s}$ matching and mismatching configurations are denoted as the P and AP states, respectively. The corresponding spin-resolved transmission spectra and edge-TMR are shown in Fig.~\ref{fig3}(c) and (d). Owing to the different edge terminations, the nanoribbon widths of the type 1 and type 2 devices are 18.1 and 20.1~\AA, respectively, which results in a larger bulk-state contribution to the transmission in the type 2 configuration than in the type 1 configuration [e.g., at $E = -0.4$~eV]. Here, the edge-TMR is defined as $\text{TMR}(E) = \left[T_\text{P}(E) - T_\text{AP}(E) \right] / T_\text{AP}(E)\times100 \% $. 

Both types of devices exhibit a quantized transmission of 2 and 0 at the Fermi level in the ON and OFF states, respectively [Fig. 3(c) and (d)]. Due to the fully insulating nature of the OFF state, extremely large TMR values of up to $10^3\%$ (type 1) and $10^7\%$ (type 2) are observed near the Fermi level, indicating promising potential for device applications. According to the calculated transport eigenstates, the ON states are dominated by spin-polarized edge states, whereas the OFF states lack conducting channels. These transport eigenstates [Figs. 3(a) and (b)] correspond to the edge states in the type 1 and 2 configurations, respectively. We also calculate the band structures of AB-stacked Cr$_2$Se$_2$O with different numbers of layers to infer the transport properties of edge-MTJs in multilayer configurations (Fig. S11) \cite{SupplementalMaterial}. The result shows that the number of conductive channels increases with the layer number. However, beyond a certain thickness, some bulk bands begin to cross the Fermi level. In this case, the device is no longer only dominated by edge conduction. Therefore, increasing the number of layers involves a trade-off between enhancing the conduction current of the edge-MTJ and maintaining transport solely through the edge states.

\begin{figure}[t]
\centering
\includegraphics[width=\linewidth]{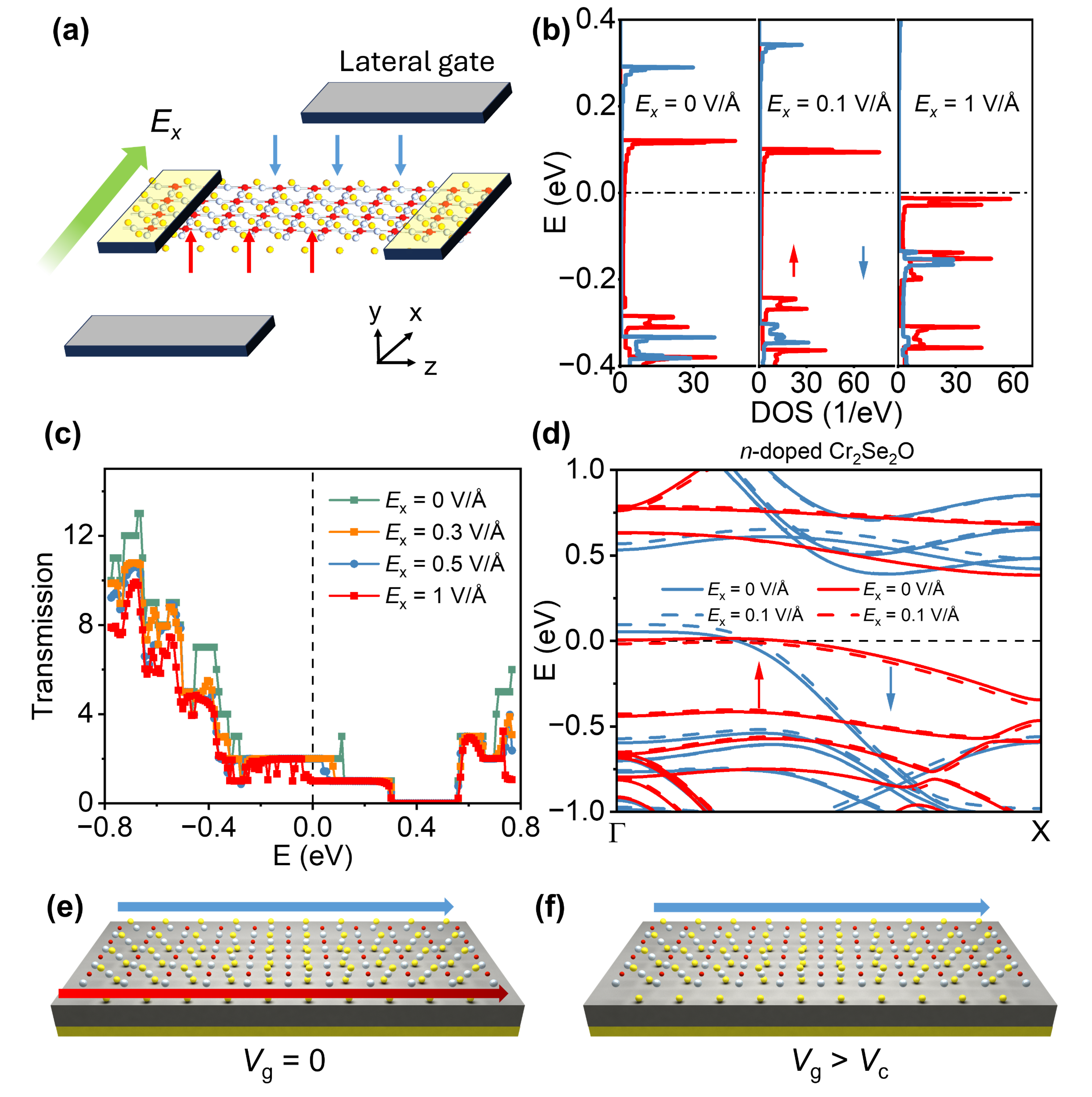} 
\caption{\label{fig4} Electric-field-tunable edge-MTJ properties. \textbf{(a)} Schematic diagram of the edge-MTJ device controlled by the lateral gate voltage. \textbf{(b)} Spin-resolved density of states in  Cr$_2$Se$_2$O nanoribbon under lateral electric field $E_x$. \textbf{(c)} Transmission spectra of edge-MTJ in the P state under different $E_x$. \textbf{(d)} Band structure of Cr$_2$Se$_2$O nanoribbon under the electron doping of 0.0225 $e$/atom, where the critical electric field for a single quantum spin channel at the Fermi level is reduced to $0.1~\mathrm{V/\mathring{A}}$. Electrostatic doping induces spin-polarized current in altermagnetic edge-MTJ. \textbf{(e)} Spin-neutral current at zero gate voltage and \textbf{(f)} spin-polarized current above the critical gate voltage. The red and blue arrows represent spin-up and spin-down, respectively.}
\end{figure}

\subsection{Electrical tuning of edge-MTJs}
For systems with $\boldsymbol{r}$-$\boldsymbol{s}$ coupling, one can effectively modulate the energy levels of different spin via the Stark effect \cite{gong2018electrically,han2024cornertronics}. We consider an in-plane electric field applied perpendicularly to the transport direction of the nanoribbon [Fig.~\ref{fig4}(a)]. The electric potential of the edge of the device near the positive/negative voltage terminal is correspondingly raised/decreased (Fig.~\ref{fig4}(b) and Fig.~S12 \cite{SupplementalMaterial}). Applying an electric field along the $x$-direction thus energetically down-shifting the spin-up edge state on one edge while energetically up-shifting the spin-down edge state on the opposite edge. When the electric field lies in the range of $0$ to $1~\mathrm{V/\mathring{A}}$, although the energy bands shift, the number of edge states crossing the Fermi level remains unchanged. When the electric field exceeds $1~\mathrm{V/\mathring{A}}$, all spin-up edge bands are shifted away from the Fermi level, leaving only a single spin-down edge states crossing the Fermi energy. Consequently, this leads to the peculiar transport window at which the transmission spectrum at the Fermi level is contributed by only 1 conduction channel carrying one spin species localizing along one edge.

We calculate the transport properties of a laterally gated field-effect device as shown in Fig.~\ref{fig4}(c) to confirm the electrically tunable fully spin-polarized quantized edge conductance. Such effect manifests as the switching of the transmission channels from 2 to 1 when the applied lateral gate field exceeds the threshold electric field of $1~\mathrm{V/\mathring{A}}$. However, it should be noted that an electric field of $1~\mathrm{V/\mathring{A}}$ is practically challenging to realize in typical device setup \cite{wang2025pressure}. One way to bypass such a large lateral gate field value is to employ electron doping in the Cr$_2$Se$_2$O channel, which can effectively lower the critical electric field to an experimentally accessible range. Such an electron doping effect can be achieved electrostatically via back gating or by using polar substrates. With an appropriate electron doping concentration of $0.0225~e$/atom, the critical field can be reduced to $0.1~\mathrm{V/\mathring{A}}$ (10 MV/cm), which is well within the experimental reach \cite{xu2024single} [Fig.~\ref{fig4}(d)]. 

Due to the spin-split nature of the edge states, only a single spin-resolved conduction channel appears at specific energies, resulting in a quantized transmission of 1. As shown in Fig.~\ref{fig4}(e) and (f), the number of conduction channels can also be changed by adjusting electrostatic doping. For example, the transmission state is switched to 1 at $E=0.2$ eV without adding lateral gate voltage or switching the Néel order [Fig.~\ref{fig4}(c)]. This doping-induced spin-polarization can be experimentally implemented via a transistor-like gating structure. Unlike the mechanisms relying on the electric-field-induced breaking of hidden spin polarization symmetry in conventional antiferromagnets \cite{lv2021electric}, the electric field here does not break the symmetry between the two magnetic sublattices. The sublattice symmetry is already intrinsically broken in altermagnets. The role of the electric field is to enhance the intrinsic asymmetry between the sublattices, enabling an electrically tunable spin-polarized current. This mechanism is fundamentally distinct from conventional approaches in antiferromagnets that rely on symmetry breaking to the hidden spin polarization \cite{zhang2014hidden}. This feature is absent in conventional antiferromagnets and highlights a novel regime where a spatially localized, nonequilibrium single-spin quantum state emerges. 

\section{Discussion and conclusion}
We note that the present device configuration differs from conventional MTJs, where transport is typically mediated by bulk tunneling across an insulating barrier. In the present case, the transport is dominated by edge states, and the switching behavior originates from the spin-dependent matching and mismatching of these edge channels. The bulk region primarily serves as a structural framework that defines the boundary conditions and supports the formation of the edge states, rather than directly contributing to the transport.

We further emphasize that the present setup represents an idealized model system aimed at elucidating the underlying physical mechanism. In particular, assumptions such as well-defined edge termination, ordered magnetic configuration, and controlled Néel switching are adopted to clearly demonstrate the edge-state-driven transport behavior. While practical realizations may involve additional complexities, we consider two representative barrier geometries, namely a hollow barrier and a step-width barrier, and verify that both can reduce the electrode--electrode coupling while preserving the switching functionality (Figs.~S13--S15) \cite{SupplementalMaterial}. These results illustrate a possible device-level implementation, while the main outcome of this work is to establish a conceptual framework for edge-state-driven spin transport and switching phenomena in altermagnets.

In conclusion, we have demonstrated an edge-state-driven spin transport mechanism in two-dimensional altermagnets, where the switching behavior originates from the spin-dependent matching and mismatching of floating edge states. This work establishes a real-space perspective of $r$–$s$ coupling and highlights the role of symmetry in enabling unconventional spin functionalities beyond momentum-space mechanisms. Our results provide a conceptual framework for exploring edge-state-based spin transport and switching phenomena in altermagnetic systems.

\begin{acknowledgments}
This work is supported by the Singapore Ministry of Education (MOE) Academic Research (AcRF) Tier 2 grant under the award number MOE-T2EP50125-0019. X.W. thank the Australian Research Council Discovery Early Career Researcher Award (Grant No.~DE240100627) for support. J. W. thanks the China Scholarship Council (CSC). J.L. thank the National Natural Science Foundation of China (Nos.~12274002 and 91964101), the Ministry of Science and Technology of China (No.~2022YFA1203904), the Fundamental Research Funds for the Central Universities, and the High-performance Computing Platform of Peking University. Y.S.A. acknowledges the support from the Kwan Im Thong Hood Cho Temple Early Career
Chair Professorship.
\end{acknowledgments}
\bibliography{apssamp}
\nocite{*}
\end{document}